# Non-spectral methods of analysis of the internal gravity waves measurements in ocean

Vitaly V. Bulatov
Institute for Problems in Mechanics
Russian Academy of Sciences
Pr.Vernadskogo 101-1, 117526 Moscow, Russia
bulatov@index-xx.ru

**Abstract.**

The paper is devoted to the presentation of the non-spectral methods of analysis of the natural measurements of the internal gravity waves in the ocean with the purpose to determine characteristics of the wave-trains composing the measured field, the forms and parameters of the ocean along the pass of these wave-trains propagation. The problem of the analysis of the data of the natural measurements of the internal waves with the purpose to separate the single wave-trains from the measured field, and on this basis to determine the characteristics of these wave-trains, to receive the information on the sources of excitation of the waves, and the information on the properties of the ocean along the pass of the waves propagation, is directly connected with the problems of the wave dynamics. The basis of the offered algorithms of the analysis is the supposition , that the measured wave field represents the sum of the plane wave-trains having the certain speed and directions of propagation. Transformation of the packages of the internal waves, which can be determined by the given methods, may testify about the passage of the wave-trains through the big-size oceanic formations, that makes it possible to remotely determine the characteristics of these formations.

## 0.Introduction

This paper is devoted to consideration of the philosophy of the space-time spectrum analysis and to the systematic presentation of the non- spectral methods of the in-situ measurements of analysis of the internal gravity waves in the ocean, which allow to determine the characteristics of the wavetrains composing the measured hydrophysical fields: of the velocity and the propagation directions, forms, as well as the parameters of the ocean along a path of propagation of these wavetrains. The problem of the data analysis of the full-scale in-situ measurements of the internal waves on the background of the big disturbances closely adjoins the problems of wave dynamics considered in the [1-8].

It is known, that the expected measured wavetrains should be broadband, that is rather short and consisting of the rather small number of oscillations, and in this situation the traditional methods of the time-space spectrum analysis can be inapplicable. We have considered the new methods of the in-situ observations allowing single out the short wavetrains for determination of the characteristic of the wavetrains, the sources of their excitation and the parameters of the ocean along the path of propagation of the wavetrains.

The basis of the offered algorithms is the supposition, that the measured wave field is the sum of the planar wavereains possessing the definite velocity and the direction of propagation. For determination of the parameters of the wavetrains it has been offered to analyze e some functionals, that allows in the case of presence in the in-situ measurements data of the stable wavetrains to determine their characteristics.

For the numerical analysis we have used the result of the measurements of the velocity components of the flow, the fields temperatures and other characteristics gained in experiment "Mezopolygon" (the tropical part of the East Atlantics), on the Western Sahara shelf and also of the experiment at the Black sea. The conducted analysis demonstrates, that really, the analysis of the offered functionals allow with the sufficient degree of accuracy to determine the lifetime of the stable wavetrains, and also their parameters. Comparison of the obtained results with the results of the usual time-space spectrum analysis of the same in-situ measurements data demonstrates the coincidence of the characteristics determined by the different measurement procedures.

The fundamental virtue of the offered methods consists, at first, that there is a possibility of the reliable analysis of the full scale data gained by means of rather small number of sensor units, while the time-space analysis superimposes the certain limitations on the layouts of the detectors in the water space, and, secondly, enables to study the transformation of the internal wavetrains along the rather long path for the purpose to determine characteristics of the large oceanic formations.

**1. Main ideas and methods of the spectral analysis of the internal gravity waves.**

The oscillating systems are widely presented in the real ocean, and the spectral description of the coupled with them random hydrophysical fields is quite natural as it is directly dealt with the intensities of oscillations at the different frequencies.

Visualization of the spectral description of such a field is pertinent to the singularities of the human perception based mainly on the visual information. We shall observe the stages of transfer of this information. The objects of the physical world surrounding us cannot be perceived directly (except for touching). The objects form the wave field propagating unrestrictedly and reaching the human eye, in which the crystalline lens performs Fourier spatial transformation translating the field into its spatial spectrum. This spectrum as the durable practice of the people demonstrates, rather well matches to the initial set of the objects. The line including the compact physical object - the infinite wave field - the compact spectrum of the field (the image of the object) supplies us with the most part of the information on the world. . Researching the oceanological field in terms of its spectrum, we reproduce the second part of the informational line-up and the transform unlimited in time and space elementary waves forming the field into the compact point components of the spectrum of the field. Thereby we put into operation our customary conceptions and the intuition, developed in the process of the long-term evolution.

In the process of any physical experiment the random field can be metered only within the limits of the restricted time and space, that is the time and the space of supervision. Repetition of the experiment yields the new realization of the field, which is also limited by time and space. Using such data it is necessary to determine the properties of the whole field - its correlation function or its spectrum.

The random character of the available implementations, or to put this other way, the limited complete set of the initial information on the field hinders to make the exact determination of the field. spectrum. On the basis of the observations it is possible to determine only the estimate of the spectrum, which always differs from the true spectrum. Its difference from the true spectrum is described statistically assuming the multiple collections and processing of the single-type initial data.

The fact, that the real physical fields do not comply with the strict mathematical determination of the random field as the aggregates of the infinite set of the infinite lengthy realizations, as a rule, is not discussed. Limitation for the sets, time and space is specially obvious in respect to the oceanologic fields, but usually it is assumed, that the theoretically necessary infinite sets are existing, although in the field measurements of the random hydrophysical fields they are accessible for us only within the limits of the restricted time and space of observations.

The present paper briefly formulates the main ideas and methods of the classic space-time analysis of the random hydrophysical fields. As it is known, the ocean can be determined as the collection of the different kind random fields - the fields of temperature, pressure, current speed, bottom surface elevation etc. From the physical point of view this not so correct determination emphasizes two singularities of the ocean - the variability of any its parameters in time and space and the random character of these variations.

The rather complete statistical description of the variability of the parameter forming the random field, at the reasonable limitations applied on the field properties is provided by the correlation function and the energy spectrum. These two nonrandom functions of the space-time shifts and the spatial-temporal frequencies are linked among themselves by the unambiguous and invertible Fourier transform. The use of this or that of them is dictated only by convenience and visualization of presentation and the further usage of the information.

The first function determines the statistical dependency between the separated by time and space measurements (readings) of some hydrophysical random field. The second function determines the distribution of the intensity of the oscillations on the frequencies, both in time and space.

In the experimental researches the justified preference is given to the spectral description, in particular, due to its proximity to the nature of the visual information. The accessible for the researcher information on the random field is contained in one or several limited in time and space realizations of the random hydrophysical field. With the help of these data it is necessary to characterize the whole field, to discover or to evaluate the spectrum of the field. Because of the random character of the available data it is impossible to give the exact description of the spectrum, but it is possible to make only its estimation. For this purpose it is necessary to fulfill two fundamentally important operations: using the Fourier transform to transfer the available data over to the frequency space, and the data are at that or other extent averaged for the statistical stabilization of the analysis result.

The estimation always differs from the true spectrum. Naturally one should seek, that this difference was minimal. The problem about the criterion of the minimum concerns the statistical properties of the estimation. The deviation of the estimation from the true spectrum can be characterized completely enough by mistakes of two types: the deviation of the concrete estimation from the mathematical expectation of the estimation value – this is the random error; - and the deviation of the mathematical expectation of the estimation value from the true spectrum – this is the systematic error.

The random mistake describes the selective variability of the results of the analysis at repetition of the experiment, the systematic mistake determines the so-called resolution of the analysis, that is the capability to reproduce separately two close by the frequency peaks of the spectrum. The mistakes can be explained as the uncertainty of the measurement of the spectral density (the random error) and the uncertainty in the position of this spectral density on the axis of the frequencies (the systematic error) and to represent it in the form of the ellipse of dispersion of the estimation around the selected true value (or the true estimation) of the spectrum.

The errors are qualitatively various, but their numerical values are linked among themselves by the relation of the uncertainties: at the fixed volume of the initial data to diminish one of the errors is possible only by increasing the other error. The ellipse of dispersion can be compressed and extended, but it is not allowed to reduce its area. The relation of the uncertainties results in ambiguity of the criterion, which might be used for the search of the best estimation.

Until recently the dominating position in the practical spectral analysis was occupied by the linear algorithms of the data processing. They were built for determination of relationships for the functions of the correlation and the spectrum. The initial data are subject to the Fourier transformation. In the analogue analyzers this operation is fulfilled by the narrow-band filters. Then the spectrum is smoothed over the frequency or averaged over the set of realizations, which ensures statistical stabilization of the estimation. The same estimation can be calculated in other succession: first it is necessary to find the estimation of the correlation function, to reject or kill its least exact values and then to make the Fourier transformation.

Before the analysis it is necessary to make the choice between the degree of the detailed elaboration of the estimation and the degree of the statistical averaging (between the systematic and random errors of the analysis). Resolution of the linear analysis is limited by the reciprocal to the time of observation. The linear estimations do not allow to determine the spectrum characteristics of the short realizations, even if there is a lot of them.

The linear algorithms impose the definite structure on the studied process, in particular, the process can be received from the white noise – the noise with the uniform frequency spectrum - using the linear filtering. At that the response of the filter is time-limited by duration of observation. The linear estimation may be considered as the result of the fitting of the indicated type model process to the initial data.

The mathematical expectation of the linear estimation of the spectrum is the convolution of the true spectrum with the spectral window. At the sufficient averaging the measured spectrum is close to its mathematical expectation and it is possible to try using the measured spectrum to retrieve the true spectrum, solving the convolution equation. This procedure allows to overcome the limitation of the resolution inherent to the linear estimations. In the general case the problem of the recovery is incorrect, has no the unique solution and requires introduction of the additional limitations or assumptions.

The second important group of the algorithms of the data processing is based on adjustment of the autoregressive processes to the initial data. Such processes (successions) originate at the reverse action on the filter with the response of the restricted duration: the initial process is fed to the inlet of the filter and on the screen outlet the white spectrum process (succession of the uncorrelated readings) is formed. The autoregressive estimations are non-linear, their statistical properties are more complicated, than those for the linear estimations. The resolution of the estimates depends on the form of the spectrum, on elevation of the resolved peaks above the medium level of the spectrum. At the short realizations it is higher than at the linear estimations of the spectrum – and that is the important distinctive feature of the autoregressive estimations, which provides their quick propagation after rather recent formation. For them the common principle of the inverse dependence between the systematic and random errors is fair, but there is no limitation on the resolution interval.

At usage of the autoregressive estimations it is necessary to make of the choice of the length of the filter or the order of the autoregression, that is analogous to the selection of the degree of the averaging in the linear algorithms. Asymptotically at unlimited lengthening of the realization the autoregressive estimations give the same result, as well as the linear estimations.

The autoregressive estimations undoubtedly are preferential: at the analysis of the short realizations, in particular, at the analysis of the non-stationary processes, the vertical cuts of the ocean; at the analysis of the field represented by a small number of the long time series; at separation of the certainly existing with similar frequencies peaks of the spectrum, etc.

Transition to the analysis of the fields is accompanied by complication of the technical side of the issue, but almost does not affect the principles of the analysis. Activity in the multidimensional space is much easier at the figurative geometrical perception of the data, procedures and the results of the analysis. The full-scale realizations of the field are metered seldom. Often it is necessary to deal with the different kind of the incomplete data about the field. The widely spread time series represent the linear sections of the fields. The special algorithm for processing the two alternately directed sections of the field allows on the base of such data to obtain the estimation of the entire of the field. The technical advantages are created by the method of the aperture synthesis applied to the data of the two meters, which in succession are taking different mutually located positions in space.

The meters quite often are moving relatively to the field, that results in the turning of the frequency coordinates, in which the analysis is carried on. The interference or antenna systems of sensors, which possess selectivity of the direction of the incoming waves, create some technical simplifications. The resonance interference systems besides kill the long-wave components of the field, that is suitable at the analysis of the less intensive shortwave components. At processing of the spatial data the autoregressive algorithms are widely used because the sites of the observation are usually small, and the linear algorithms do not provide the acceptable resolution on the spatial frequencies. The statistical stabilization of such estimations is ensured by averaging of the time intervals or the time frequencies.

The enumerated two types of algorithms - the linear algorithm and the autoregressive algorithm - do not embrace all possible methods of the data processing for the purpose to determine the data spectrum. Recently other model spectrums and other approaches to the problem solution are discussed and researched. In particular, in the case of shortage of the initial data use the low-parametric models giving the approximate idea of the spectrum.

The choice of the type and the particular parameters of the algorithm is the problem individual by nature and the main by importance for the practical spectral analysis. The choice depends on the volume of the initial data and the purpose of the research, on the expected form of the spectrum, on additional information about the spectrum, on the desirable degree of the data generalization, etc. The choice may contain one or several trial calculations with the subsequent refinement of the parameters. The choice is realized in the form of the preset resolution of the analysis, the number of degrees of freedom of the estimation, in the form of selection of the definite model of the spectrum or in other form.

Although the argumentation of the choice can be well-grounded, the complete description of the spectrum according to the available data generally provides for several different its estimations, which differently present the spectrum structure. The different results of the analysis should be perceived as the different images of one and the same spectrum. Within the limits of the experiment with the restricted data volume there is and acting the same principle of complementarity, as in the other cases of the description of the phenomena and the objects having the probabilistic nature.

The spectral description is remarkable for its surprising universality and effectiveness. The understanding of the basic relationships and limitations saves the experimenter from the unjustified hopes and so unjustified scepticism in relation to the results of the analysis.

Any physical, chemical, biological parameters of the water stratum of the ocean – the temperature, saltiness, flow rate, the surface level, transparency, contents of oxygen, concentration of inhabitants and others - depend on each other and on the external actions and are randomly changing in time and in space forming the random fields. The wind disturbances represent the field of the elevated surface with respect to the quiet level; the turbulent movements of water cause the random fields of temperature, pressure, speeds. As it was noted above, for description of the variability of the parameter forming the random field, they mostly use the correlation function and the energy spectrum of the field.

The correlation function is the mathematical expectation (the average value obtained as the result of the set of realizations and marked by the character $E$) of the product of the shifted in time and space values of the field $\xi(\mathbf{X})$

$$B(\Delta \mathbf{X}) = E[\xi(\mathbf{X})\xi(\mathbf{X} + \Delta \mathbf{X})].$$

Here, as everywhere, where it does not result in the misunderstandings, for reduction of the writing the vectors $\mathbf{X}$ and $\Delta \mathbf{X}$ include all space coordinates and the time. At that usually the vectors and their components are not divided according to their designations

$$\mathbf{X} = (x, y, z, t); \qquad \Delta \mathbf{X} = (\Delta x, \Delta y, \Delta z, \tau).$$

As a rule the scalar fields – the fields of the temperatures, the surface elevation, etc - are usually considered in practice of the oceanologic in-situ measurements and at their spectral processing. Usually it is supposed, that the considered fields are the stationary and uniform fields, which static properties (any average indices) do not vary neither in time, nor in space, therefore the correlation function does not depend on $\mathbf{X}$. The stationarity and the uniformity of the fields as a rule are linked with the ergodicity of the field, which allows to substitute the averaging o f the set of the realizations for the practically more accessible averaging in time and in space:

$$B(\Delta \mathbf{X}) = \lim_{\Pi \to \infty} \frac{1}{\Pi} \int_{\Pi} \xi(\mathbf{X})\xi(\mathbf{X} + \Delta \mathbf{X}) d\mathbf{X},$$

where $\Pi$ is the volume of the space, in which the realization is conducted.

The correlation function possesses a central symmetry

$$B(\Delta \mathbf{X}) = B(-\Delta \mathbf{X}),$$

and it is nonnegative in zero

$$B(0) = a^2 + \sigma^2 > 0,$$

where $a$ is the average value of the field, $\sigma^2$ is the dispersion;
its principal maximum is disposed in the coordinates origin

$$B(0) \geq B(\Delta \mathbf{X}).$$

The correlation function is the positively definite function, i.e. for any real $b_i$ $(i = 1, N)$ the following inequality is fair:

$$\sum_{i,j=1}^{N} B(\tau_j - \tau_i) b_i b_j \geq 0.$$

The infinitesimal of the correlation function at some shift $\Delta \mathbf{X}$ means, that on this interval on the average the realizations are essentially changing their current value. At expansion of the shift the correlation function is monotonically or with oscillations decreasing.

The function conjugated by Fourier with the correlation function is the energetic spectrum of the field $g(\mathbf{k})$:

$$g(\mathbf{k}) = \int B(\Delta \mathbf{X}) \exp(-i2\pi \mathbf{k}\Delta \mathbf{X}) d\Delta \mathbf{X}$$ - Fourier forward transform;

$$B(\Delta \mathbf{X}) = \int g(\mathbf{k}) \exp(-i2\pi \mathbf{k}\Delta \mathbf{X}) d\mathbf{k}$$ - Fourier inverse transform;

where $\mathbf{k} = (k_x, k_y, k_z, f)$ - the vector of frequency including the time frequency $f$ and the space cyclical frequencies, or the components of the wave vector $k_x, k_y, k_z$: $\mathbf{k}\Delta \mathbf{X} = k_x \Delta x + k_y \Delta y + k_z \Delta z + f\tau)$ - scalar product of vectors $\mathbf{k}$ and $\Delta \mathbf{X}$.

Here and in the further the limits of integration are not indicated, if they spread from $-\infty$ up to $+\infty$.

As a rule, the term "frequency" can be applied both to the time and the space oscillations, so that the time period and the wavelength are equal to the inverse value from the time frequency and the modulus of the space frequency. Function $g(\mathbf{k})$ is also called the energetic spectral density (the Fourier density of dispersion) – this is the most accurate, but the bulky name - or it is called just the spectrum.

Dimension of the spectrum is a square of the random value divided by the unit volume of the frequency space. The Fourier density cannot receive the negative values (this property follows from the positive determinancy of the correlation function):

$$g(\mathbf{k}) \geq 0$$

and possesses the central symmetry owing to the central symmetry of the correlation function

$$g(\mathbf{k}) = g(-\mathbf{k}).$$

Considering this property, the spectrum is usually figured and discussed in one frequency half-space, for example, only for the positive time frequencies or only for the negative time frequencies.

The averaged from the set of realizations intensity of oscillations, which frequencies lie in the small volume (interval) $\Delta \mathbf{k}$ around of the frequency $\mathbf{k}$, is equal to the double product of the Fourier density by the volume $\Delta \mathbf{k}$:

$$2g(\mathbf{k})\Delta \mathbf{k}$$

where the coefficient 2 allows for the Fourier density in the second frequency half-space. The volume under the whole spectrum is equal to the total intensity (dispersion) of the field, if the average value of the field is equal to zero:

$$\sigma^2 = \int_{-\infty}^{+\infty} g(\mathbf{k})d\mathbf{k} = 2 \int_{0}^{+\infty} g(\mathbf{k})d\mathbf{k}.$$

The physical idea of the energetic spectrum is linked to разложением поля of the field into its sinusoidal components: the spectrum describes, what contribution into the total dispersion of the field is introduced by the components with the different frequencies, that is it presents the distribution of the intensity of the oscillation at each frequency.

To the correlation function, which is monotonically decreasing with the increase of the shift, corresponds the spectrum, which is monotonically decreasing by frequency. The central frequency of such a spectrum determined as the center of gravity of the spectrum in the frequency half-space characterizes the width of the spectrum. Such spectrum and the field are called broadband. Compression of the correlation function along any component of the spatial shift vector results in expansion of the spectrum along the corresponding component of the space frequency vector.

For the oscillating correlation function the spectrum looks like the narrow peak removed from the home of the frequency coordinates. Such a spectrum and the field are named the narrowband. For the spatial-time cosinusoid the spectrum has the form of two delta-functions.

The limiting cases of the monotone correlation function are the delta-function and the constant function:

$$B(\Delta \mathbf{X}) = \delta(\Delta \mathbf{X}) \quad B(\Delta \mathbf{X}) = \text{const}$$    The limiting case of the oscillating function is the cosinusoid:

$$B(\Delta \mathbf{X}) = \cos 2\pi \mathbf{k}_0 \Delta \mathbf{X}$$

The corresponding spectrums look like:

$$g(\mathbf{k}) = \text{const} \quad g(\mathbf{k}) = \delta(\mathbf{k}) \quad g(\mathbf{k}) = \frac{1}{2}[\delta(\mathbf{k} + \mathbf{k}_0) + \delta(\mathbf{k} - \mathbf{k}_0)]$$

Quite often one works with complex random processes and fields, each reading of which is a complex number. The complex presentation is suitable and economical at the description of the narrowband oscillations. These oscillations are replaced by the enveloping curve, which is represented in the form of the complex readings consisting of the modulus and the phase of oscillations. The readings of the enveloping curve can be taken with a time step equal to the reverse value from the width of the peak of the spectrum, whereas the readings of the process itself should be taken more often, than by two per the oscillation period. The gain in the number of the readings is equal to the degree of the narrowing of the band of oscillations - the ratio of the current frequency to the width of the peak of the spectrum.

The narrow-band oscillations are widely presented in the ocean. The spatial analysis usually is conducted separately for oscillations with the fixed time frequencies, presenting data in the form of the complex readings of the amplitude-time spectrum.

The acoustic and radio signals usually are the narrow band signals and are used at the non-contact measurements of the oceanologic parameters. At determination of the correlation function the second shifted reading of the complex field is expedient to take with the complex conjugation (marked by a sprocket):

$$B(\Delta \mathbf{X}) = E[\xi(\mathbf{X}) \xi^*(\mathbf{X} + \Delta \mathbf{X})]$$

This ensures the real nature of the energetic spectrum. For the rest all the defining relations for the correlation function and the spectrum remain valid. The spectrum of the complex process is non-symmetric concerning the zero frequency. The spectral expansion can be applied directly to the realization of the field $\xi_X(\mathbf{X})$ limited by the volume $\Pi$

$$Q_X(\mathbf{k}) = \int \xi_X(\mathbf{X}) \exp(-i2\pi \mathbf{k}\mathbf{X}) d\mathbf{X}.$$

The complex function $Q_X(\mathbf{k})$ describes the spectral density of the amplitudes of the sinusoidal and cosine components forming the realization, and is called the amplitude spectrum of realization. The amplitude spectrum can be determined using Fourier forward transformation from realization:

$$Q_X(\mathbf{k}) = \int \xi_X(\mathbf{X}) \exp(-i2\pi \mathbf{k}\mathbf{X}) d\mathbf{X}.$$

In the assemblage of the spatially limited realizations the modulus of the amplitude spectrum is rather stable, its mathematical expectation value is nonzero, and the phase of the spectrum is accidental and distributed uniformly within the limits from 0 up to $2\pi$. Transition from the spatial coordinates to the frequency coordinates results in replacement of realization of the random function $\xi_X(\mathbf{X})$ for realization of the random complex function $Q_X(\mathbf{k})$.

At the volume X unlimited expansion any neighboring values of the amplitude spectrum turn out to be mutually uncorrelated, and the mathematical expectation of the square of the spectrum modulus seeks to the fixed nonnegative value depending on the frequency $\mathbf{k}$. For the conditionally introduced amplitude spectrum $A(\mathbf{k})$ of the unlimited in space realization the following ratio is fair:

$$E[A(\mathbf{k}) A^*(\mathbf{k} + \Delta \mathbf{k})] = g(\mathbf{k}) \delta(\Delta \mathbf{k}),$$

where the sprocket indicates the complex conjugation; $\delta(\mathbf{k})$ is a many-dimensional delta-function; $g(\mathbf{k})$ is the energy spectrum of the field. This ratio is relevant in the practice because it links the energy spectrum of the field with the Fourier transform from realization of the field and the part of the linear algorithms for the energy spectrum estimation and the corresponding instrumentation are based on it.

The correlation function and the energy spectrum contain the same information on the spatial-temporal properties of the fields of the internal gravity waves, and in this sense the descriptions of the correlation and spectral fields are equivalent, however in the physical space the information is expressed differently, than in the frequency space.

However, if the measured fields of the internal gravity waves contain the significant noise components, then it is more suitable to use other methods of the analysis of the in-situ measurements different from the spectral methods. These methods are mainly based on such properties of the real packages of the internal gravity waves as, for example, inharmoniousness, and the following sections of this research are devoted to their systematic presentation.

## 2. Non-spectral methods of the analysis of the measurements of the internal gravity waves in the Western Sahara shelf region.

This Section presents the results of the measurements of the internal gravity waves obtained during the 52-nd voyage of "Michael Lomonosov" research ship. Presented are the methods of processing of the in-situ measurements, which allow to spread out the measured field of the internal gravity waves to the sum of the separate wave trainы and considered the results of the measurements of the internal gravity waves performed during the 52-nd voyage of "Michael Lomonosov" research ship, and also the methods of their processing, which unlike the known methods of the time-space spectral analysis, and also other methods of extraction of the signals on the background of the accidental hydrophysical fields, allow to present the measured field of the internal gravity waves in the form of the sum of the relatively small number of the separate wave trains, each of which is characterized by the certain parameters: the velocity, the propagation direction and the form{shape}of the wave.

Measurements of the temperature, the components of the horizontal velocity with the vector-averaging were exercised by the "Vostok" measuring complex developed in MGI (the Ukraine Academy of Sciences). The instruments mounted on the various depths were arranged on the four anchored buoy stations (internal buoyancy), forming the space antenna-rhomb with the side of about 220 meters and the smaller apex angle of 80˚.

Observations were conducted in region of the Western Sahara shelf in the point with coordinates 20°55 ' N, 17°34 ' W, the observations recording was led during 4 days from March 20th till March 24th, 1990 with the interval of 15 seconds.

At the primary processing of the in-situ observations data they used the time window allowing in further to consider the function with the spectrum, lying being in the interval typical for the internal gravity waves

$$S(t) = R(t) - \frac{1}{2T_2} \int_{t-T_2}^{t+T_2} R(\rho)d\rho \equiv R(t) - E(t).$$

$$R(t) = \frac{1}{2T_1} \int_{t-T_1}^{t+T_1} I(\rho)d\rho,$$

where I(t) is the initial signal record made by the sensor; $T_1 = 100$ seconds, $T_2 = 1$ hour. The Fig. 2.1 presents the typical record of one of the instruments (the meridianal velocity) of three days duration: the solid line is S (t), the dashed line is the function *E (t),* describing the slow variations of the meridional velocity (the tidal stream).. From the Figure one can see, that the internal gravity waves are propagating predominantly by groups, and the oscillations inside the groups are of еру quasi-harmonic nature with the oscillation period - $T^*$ of the order of ten minutes, at the same time there are rather the high-frequency oscillations outside the groups, as a rule of the smaller intensity.

The analysis of the measurements conducted within 4 day of the observations demonstrates, that one can see the rather distinct daily recurrence of the temperature variations as well as the components of the horizontal velocity with the big amplitude possibly interlinked with the tidal stream. It is also necessary to mark the coincidences of these time intervals, when the rather big amplitudes of the oscillation of the horizontal velocity components and the temperatures were recorded. Further just these time intervals are considered.

The basis of the following analysis is the supposition, that the measured field represents the sum of the rather small number M of the wavetrains of the plane waves and some noise addition Q:

$$S(t,x,y,z) = \sum_{m=1}^{M} \psi(z) f_m(c_m t - x \cos\varphi_m - y \sin\varphi_m) + Q, \qquad (2.1)$$

Here S is any of the being measured values (temperature, horizontal components of velocity); $\psi_m$ - describes the distributions of the field of the m-th "wavetrain" on depth; $f_m$ - describes the shape of the m-th "wavetrain"; $c_m$, $\varphi_m$ - is the velocity and the propagation direction of the m-th "wavetrain" as a whole, i.e. $c_m$ is the group velocity of this "wavetrain".

Then from (2.1) follows, that observations of the i-th sensor ($i = 1, 2, ..., P$) $S_i$ look like

$$S_i = \sum_{m=1}^{M} \psi_m(z_i) f_m(c_m t - x_i \cos\varphi_m - y_i \sin\varphi_m) + Q_i ,$$

where $(x_i, y_i, z_i)$ are the coordinates of the i-th sensor unit; $Q_i$ is the irregular noise component of the i-th sensor unit. Here $M, c_m, \varphi_m$ and the functions $\psi_m$, $f_m$ are the subject to determination. At that the criterion of the fact, that the measured field is really represented in the form of (2.1) is the smallness of the $Q_i/S_i$ ratio.

The procedure of expansion of the wave field on the sum (2.1) consists of three stages: determination of the times of delay of each "wavetrain" at their movement from one sensor unit to another; determination from the times of the delay of the velocity and the direction of propagation of each "wavetrain", and at last, determination of the wave profile. The formulated problem makes sense, only if M number of the wave trains is essentially less, than P number of the sensor units, i.e. if P readings of $S_i$ sensor units one can manage with accuracy of up to the noise addition $Q_i$ to describe using the essentially smaller M number of $f_m$ functions.

The velocity and the direction of each plane wave propagation may be unambiguously determined, if the time delays readings of this wave from two sensor units with respect to the third sensor unit are known; all sensor units should be arranged on the same depth. For determination of times of delay they use two modes. The first mode consists in in the research of the correlation function

$$K_{ij}(t,\tau) = \int_{t-T/2}^{t+T/2} S_i^*(\rho) S_j^*(\rho + \tau) d\rho ,$$

$$S_i^* = S_i - \frac{1}{T} \int_{t-T/2}^{t+T/2} S_i(\rho) d\rho ,$$

where $S_i$ are the readings of the i-th sensor unit observations; T is the typical lifetime of the stable wavetrain of the internal gravity waves. As it is possible to suppose, the value of T should be at least of not less than the typical period of oscillations $T^*$, equal, as it was mentioned above, to ten and more minutes. If the considered wave field consists of one "wavetrain", then it is obvious, that the correlation function will have the maximum value at

$$\tau_{ij}^1 = ((x_j - x_i)\cos\varphi_1 + (y_j - y_i)\sin\varphi_1)/c_1$$
$$i \neq j, \quad i,j = 1, 2, 3, 4. \qquad (2.2)$$

In the case, if the measured field is the sum of several plane waves, then each "wavetrain" with the certain time delay $\tau_{ij}^m$, as it is possible to expect, will have its own local maximum of the correlation function. Therefore further we shall be interested in the local maxima of the function $K_{ij}(t,\tau)$, which is low-variable during the time period. The conducted numerical calculations demonstrate, that mainly the functions $K_{ij}(t,\tau)$ have only one explicitly expressed stationary maximum. On Fig.2.2 there is the typical view of the meridional lines of the function $K(t,\tau)$ calculated according to the readings of two sensor units (the meridianal velocity), the beginning - 0 o'clock, March 22, 1990, T =1 hour

From Fig. 2.2 one can see, that there are time intervals of no less than 2 hours, when exists only one stationary maximum of the correlation function. At that it is necessary to note, that these time intervals with the stationary maximum of the correlation function (6 - 9, 18 - 20 hours on March, 22nd, 1990) coincide with the time intervals noted above, when the quasi-harmonic oscillations with the rather big amplitude are watched. The second mode of determination of the time delay applicable only for the spatial arrangement of the sensor units in the form of the parallelogram, consists in the following: supposition, that the wave field consists only of two wave trains and , it is possible to compose the linear combination of the functions $S_i$, in the form of ( $Q_i = 0$ )

$$L(t,p,q) = -S_1(t+p+q) + S_2(t+p) + S_3(t+q) - S_4(t)$$

which identically becomes zero at any value of t, if the values of ( $p,q$ ) coincide with two pairs of the numbers ( $\tau_{12}^1, \tau_{13}^2$ ), ( $\tau_{12}^2, \tau_{13}^1$ )determining the times of delays of two waves recorded on two sensor units. It should be noted, that the analogous, but only more cumbersome linear combination of the readings of the sensor units may be compiled also for the arbitrary spatial antenna consisting of three sensor units, but also in this case one may determine of no more than two wavetrais. Further we shall consider the functional

$$F(t,p,q) = \int_{t-T/2}^{t+T/2} L^2(\rho,p,q)d\rho ,$$

where T - as well as earlier is the typical lifetime of the stable wavetrains of the internal gravity waves. It is evident, that the points of minimum of this functional determine the corresponding times of delay of two wavetrains. As the results of the numerical calculation demonstrate there are two typical structures of the disposition of the minimum of the functional $F(t,p,q)$ on the plane ( $p,q$ ). If in the examined signal there is domination of only one wavetrain, then the level lines near to the minimum of the functional $F(t,p,q)$ have the form of the "cross"; the coordinates of the cross points of the "cross" and the corresponding axes determine the times ofdelays of this wavetrain. The Fig.2.3,a presents the level lines of the functional (meridianal velocity) for 18 hours on March, 22nd, 1990; the Fig.2.3,b presents the level lines of the functional $F(t,p,q)$ for 1 hour 30 minutes on March, 22nd, 1990. The conducted numerical calculationsdemonstrate, that the presence of the second internal gravity wave is determined more seldom and only in the case of diminution of the value of T , that proves the rather small life duration of the second "wavetrain". Besides such a pattern of the level lines is not stationary enough.

The obtained values of the time delays $\tau_{ij}^1$ of one wave have allowed to determine the velocity and direction of propagation of the given wavetrain of the internal gravity waves from the overdetermined system of the equations (2.2). The overdetermination of the system (2.2) allows to determine these values by the different methods, that enables to examine the reliability of the obtained results.

This way obtained the value of the propagation velocity of the "wavetrain", that is the group velocity, has practically coincided with the maximum group velocity of propagation of the long internal gravity waves,, which was calculated by the Brent-Вяйсяля frequency distrubutin and is equal to $c_1 = 0.23$ m/s. It should be also to noted, that this separable and mainly stable "wavetrain" of the internal gravity waves comparatively little (within the limits of 20° ) changed its direction of propagation during the whole period of observation , that can indicate the uniform and rather motionless source of the disturbance of the internal gravity waves. At the same time the velocity and the direction of propagation of the second wave train selected on the background of the first wavetrain have often varied.

As it is known, the field of the internal gravity waves at the great lengths of the waves is almost longitudinal. Physically it means, that oscillations of the liquid particles should exist along the direction of propagation of the "wavetrain". Check of this property allows to gain the additionalconformation of the gained results, since during the pass of only one planar "wavetrain" all the oscillations of the particles practically should take place in the horizontal plane and in the direction of propagation of this "wavetrain". Therefore we shall represent further the vector of the level velocity **u** in the form of $\mathbf{u} = \mathbf{u_1} + \mathbf{u_2}$, where $\mathbf{u_1}$ is the component of the vector along the direction of the wave propagation, $\mathbf{u_2}$ is the component normal to the direction of the wave propagation.. Fig. 2.4,a presents the modulus of the component $|\mathbf{u}|$, the Fig.2.4,b presents the modulus of the component $|\mathbf{u_1}|$, the Fig.2.4,c presents the modulus of the component $|\mathbf{u_2}|$, the beginning - at 12 clocks, March 22, 1990. Really, as it is demonstrated by the given expansion of the measured horizontal velocity in direction of propagation of the wavetrain (the value of $\varphi_1$ is determined from (2.2)), during its passage mainly all oscillations happen along this direction, and deflections of the liquid particles in this direction more than five times exceed the deflections of the particles in the direction normal to the direction of propagation of this "wavetrain".

In conclusion of the present paragraph it is necessary to pay attention to the problem of determination of the shape of the plane wave, i.e. in this case - to the finding of the function $f_m$ from the series of (2.1). The true shape of the " wavetrain" can be gained by summation of the functions $S_i$ for the sensor units arranged on the same depth,

$$f_m \approx \frac{1}{P} \sum_{i=1}^{P} S_i(t + \tau_{ij}^m), \qquad i,j = 1,2,...,P,$$

where $\tau_{ij}^m$ - the corresponding times of the delays, because it is possible to expect, that the sum of the functions $\sum_i Q_i$ will be much smaller, than the base signal repeated mainly by all the sensor units. Really, as one can see from the Fig.2.5, which represents the function $S_4$ (the meridianal velocity, the beginning - 12 clocks on March, 22, 1990 (Fig. 2.5a), function $f_1$ (Fig..2.5,b) and the difference between them (Fig..2.5, c) for the considered in this paragraph in-situ data the shape of the plane wave during the passage of only one " wavetrain" is mainly iterates by all the sensor units and is rather feebly distorted by the random noise interferences.

**3. Non-spectral and the spectral researches of propagation of the tidal internal gravity waves on the example of the experiment – «Mezopolygon» (the tropical part of the Eastern Atlantic)**

This section presents the methods of the analysis of the accidental hydrophysical fields, which allow to determine the velocity and the direction of propagation of the plane wavetrains composing the measured wave field and sets forth the series of the numerical results illustrating the application of this method on the example of the data gained in the indicated experiment "Mezopolygon". The wavetrain field is represented in the form of the sum of the plane waves, each of which is characterized by own velocity and direction of propagation. At that we firstly present the new non-spectral method of the analysis of the hydrophysical fields disturbed by the accidental interferences, and secondly, we set forth the results of application of this method in the comparison with the results of the traditional methods of the spectral processing of the data gained in the experiment «Mezopolygon» (the tropical part of the Eastern Atlantic).

*Non-spectral data processing algorithm.* The offered non-spectral method presents the development of the approaches to the analysis of the measurements of the accidental hydrophysical fields, which have been set forth in the section 2 and representing some analog of the "wavelet transform" methods. Further we shall suppose, that the measured hydrophysical field represents the sum of the plane wavetrains:

$$F(t,x,y,z) = \sum_{m=1}^{M} \Psi_m(z) f_m(x\cos\varphi_m + y\sin\varphi_m - c_m t),$$

where M - the total number of the wavetrains composing the measured field; F - any of the measurands ( the temperature, the velocity components, etc.); $\Psi_m(z)$ - the function describing the distribution of the i-th "wavetrain" depthward; $f_m$ - describes the form of the "wave train"; $c_m$, $\varphi_m$ - the velocity and the direction of propagation of the m-th "wavetrain" as a whole ( $c_m$ - the group velocity). It is necessary to underline, that, any limtations on the particular kind of the functions $f_m$ are not superimposed. Then observations of i-th sensor unit arranged in the point with ccoordinates ($x_i, y_i, z_i$), will look like

$$F_i = \sum_{m=1}^{M} \Psi_m(z_i) f_m(x_i \cos\varphi_m + y_i \sin\varphi_m - c_m t) + Q_i, \qquad (3.1)$$

where $Q_i$ - the accidental noise component of the readings of the i-th sensor unit. Presentation of the expected on the sensor units signals in rather simple form (3.1) is based on the results gained at solution of a series problems about propagation of the different hydrophysical (ultrasonic, gravity, etc.) fields in the non-uniform mediums.

Besides well enough developedd methods of the asymptotic analysis exploring the evolution of the locally harmonic wavetrains in the non-uniform and non-stationary mediums, there are the methods, allowing to consider propagation of superposition of the harmonic waves, that is to reduce (with a sufficient degree of accuracy) an integral representation of the wave field to some final sum of the items, to which the previous chapters of the present monography have been devoted.

As it is shown in the [1-8] each derivable item is the asymptotics with respect to any parameter of the integral representation of the solution taking into account the real variability of the parameters of the medium. The phase functions of these items (Fourier integrals, Peircy integrals, Airy and Bessel functions, etc.) allow to describe the local singularities of the dispersive dependences (the local extremes of the dispersive curves, the points of their reconnections, etc.), which make the basic contribution into the integral representation of the waver field.

It is obvious, that in the representation (3.1) the phase functions are $f_m$ functions. At that the structure of the parameter of the phase functions can look like

$$\alpha(x,y,t)[x\cos\varphi(x,y,t) + y\sin\varphi(x,y,t) - c(x,y,t)],$$

where functions $\alpha, \varphi, c$ are the slowly varying (on the scale of variation of the phase function $f_m$) functions x, y, t depending, as a rule, on the several first coefficients of Taylor expansions of the dispersive curves in the corresponding critical points, and describing the dispersive effects taking into consideration the real properties of the medium. In this case the functions $c(x,y,t)$ are describing the velocity of propagation of the wavetrains "as a whole". Construction of such asymptotic forms allows to avoid cumbersome numerical calculation of the exact solutions, and in addition enables to conduct the qualitative analysis of the basic physical effects connected with particular aspect of the considered problems.

For example, according to the made estimations, for the real parameters of the stratified mediums, the expected of the internal waves ( Airy or Fresnel wavetrains) can be the broad-band wavetrains, rather short wavetrains and consist of the rather small number of oscillations. In this case conventional methods of the spectral time-multiplexed analysis can turn out inapplicable.

In Section 5.2 the velocity and the direction of propagation of the plane wavetrains were determined by study of the inconvertible stationary maxima of the correlation functions composed on basis of the readings of the different sensor units arranged on the same depth for the purpose of determination of the time delays of the definite "wavetrain" recorded by these sensor units. However using such method of determination of the characteristics of the wavetrains it is possible to analyse only the readings of two different sensor units or four sensor units arranged in the form of the spatial antenna-rhombus. This circumstance does not allow to use this method in the case of presence of the big number of the sensor units, which are not having the rather correct spatial disposition in the horizontal plane, and accordingly to analyse more effectively the readings of the significant number of the sensor units.

Before transition to the presentation of the offered method and the selection of the functional, by means of which the parameters of the wavetrains will be determined, we shall consider some leading parameters. Further we shall consider the readings of the sensor units arranged on the same depth. Let the m-th spreading "wavetrain" is recorded by the majority of N analysable sensor units. The time delay of the m-th "twaverain" through any of the i-j –the pair of the sensor units is univalently determined as:

$$\tau_{ij}^m = ((x_j - x_i)\cos\varphi_m + (y_j - y_i)\sin\varphi_m)/c_m,$$
$$i, j = 1, 2, ..., N, \quad i \neq j.$$

Then, supposing the smallness of the signal (function $f_m$) deformation at the pass of the m-th "wavetrain" through N sensor units, it is possible to expect, that the sum $\frac{1}{N}\sum_{i=1}^{N} F_i(t + \tau_{ij}^m)$ will be in the certain degree is close to the function $f_m$. As the functional, which can describe the measure of coincidence of these functions, we shall consider the following functional:

$$D(t, c, \varphi) = A(t, c, \varphi)/B(t, c, \varphi),$$

$$A(t, c, \varphi) = \int_{t}^{t+T}\left(\frac{1}{N}\sum_{i=1}^{N} F_i(u+\tau)\right)^2 du,$$

$$B(t, c, \varphi) = \int_{t}^{t+T}\left(\frac{1}{N}\sum_{i=1}^{N} F_i^2(u+\tau)\right) du,$$

$$\tau_i = [(x_0 - x_i)\cos\varphi + (y_0 - y_i)\sin\varphi]/c.$$

Here ($x_0, y_0$) are the coordinates of the sensor unit, in respect to which the time delays are calculated; T is the averaged time, which does not exceed the expected lifetime of the inconvertible "wavetrain"; $N \geq 3$ is the general number of the analysable sensor units arranged on the same depth. According to the supposition, the lifetime of the stable "wavetrain" and correspondingly epy constancy of the function $f_m$ should exceed the characteristic periods of oscillations of the studied hydrophysical fields, and the values of T should be, at least, more than the characteristic cycles of the studied signals.

Further we shall consider the behavior of the minimums of the functional $D(t, c, \varphi)$ - subject to the variables $c, \varphi$. Proceeding from the structure of this functional it is obviously, that if $D(t, c, \varphi)$ is close to zero at any values of $c_1 = c_0$, $\varphi = \varphi_0$, then it means, that all N sensor units during the time interval ($t, t + T$) registered the different random signals.

If the values of $D(t, c, \varphi)$ are close to unit at some values of $c_1 = c_0$, $\varphi = \varphi_0$ it can mean, that the "wavetrain" having the velocity $c_0$ and direction $\varphi_0$ passes through the majority of the sensor units during the time interval ($t, t + T$).

If several wavetrains transits through the system of the sensor units with the velocity and directions $(c_m, \varphi_m)$ $(m = 1, 2, ..., N)$, then it is possible to expect, that the function $D(t, c, \varphi)$ will have N local maxima $c = c_m$ $\varphi = \varphi_m$. There can be also the local maxima non-corresponding to any wavetrain. The criterion distinguishing the false maxima from true maxima is positive stability of the maximum. If the function $D(t, c, \varphi)$ has as the function $c, \varphi$ the fixed or the low shifting maximum at all t from some interval $t_0 < t < t_1$ (with the length of the interval $t_1 - t_0$ rather big as compared with the period of T averaging), then it is natural to suppose, that in this interval the "wavetrain" is propagating. The false maxima fading out at the relatively small changes of t do not correspond to any "wavetrain"

*Experiment "Mezopolygon" and the spectral analysis of the data of the measurements.* Before to describing the results of the above presented algorithm to the data of the experiment "Mezopolygon", we shall set forth the necessary information on this experiment. On the buoy stations arranged in the tropical part of of the Eastern Atlantic, where the current velocity and the temperature were measured (by the separate self-recording units). The Fig. 3.1 demonstrates the layout of the buoys, which data have been used for processing. The discreteness of the analysable field measurements compounded one hour. The predominant wave process during the experiment "Mezopolygon" was presented by the tidal internal waves with 12.4 hours phase of time and the wavelength of the order of 60-80 km (the phase velocity - about 1.5 m\s). It is known, that the internal rising tide inherits from the barotropic tide the fortnightly property (14.76 day) of the variability of amplitudes, which is bound to the beatings of the oscillations of the waves $M_2$ (the phase of 12.4 hours) and $S_2$ ( the phase of 12.0 hours). These beatings are presented on the Fig. 3.2, where the data are subjected to the band filtering for extraction of the semidiurnal period of the oscillations and for suppression of the low-frequency synoptic component and the small-scale noise. The interval of some days, when amplitudes of the waves are maximal, may be considered as the "wavetrain" of the internal tidal waves.

Generation of the internal tide happens as a result of interaction of the flows of the barotropic tide inflow with the irregularities of the ocean bottom. The currents flowing past the slope landforms of the bottom gain the vertical components and periodically displace the isolines of density in the vertical direction. It leads to the semidiurnal oscillations of the isolines of density and as a result of it to formation of the internal tidal wave. In the area of the experiment - «Mezopolygon» on the bottom of the ocean there are a lot of the submarine hills of 500-1000 m high. Such hills are arranged in each 10-20 miles. The non-uniformity of the bottom topography, which is present by the submarine low hills, is the analog of "grater", above which the barotropic inflow line is bent. Therefore, the internal inflow is excited directly in the area of the measurements. It distinguishes the area from the majority of others, where the waves come from submarine ridges or from continental slopes.

The big number of the buoy stations, with the small spacing interval between them gave the excellent opportunity to calculate the time-multiplexed spectrum of the semidiurnal internal waves. Calculation of the spectrums was conducted by means of the antenna method offered by Barber. The typical result of the calculation of the spectrum for the temperature variations on the horizon of 200 m is presented on the Fig. 3.3. This Figure shows in the coordinates of the horizontal wave numbers the isolines of the peak of the spectral density corresponding to 90, 60 and 30 % of the maximum value, which corresponds to propagation of the semidiurnal internal wave in the North-north-east direction. The wavelength is about 70 km. The spectrums were calculated both for the array of measurements on all the buoys during all the time of measurements, and for the separate groups of buoys and during a part period of the time of measurements.

All the calculations show, that on the spectrum there is one narrow precisely expressed spike on the wavelength in the range of 60-80 km with the dominanting direction of the waves propagation towards the North . Depending on selection of the group of the buoys and the time span for calculation of the spectrum the propagation direction varies within the limits of the 45˚ sector at the general direction towards the North. Such results of the calculation show, that the internal waves observed on "Mezopolygon" by their scales correspond to the second mode of the oscillations. The independent calculations using the different groups of the buoy stations for two sequential time intervals in April and May of 1985 have not gained any essential differences in the results, that testifies about the quasi-stationarity of the observed system of the waves.

The temporal variability of the quasi-stationarity of the observed system of the oscillations has been researched in the region of the Range-70, which is close to the previous area of thew measurements. Both experiments were conducted in the northern part of the basin of Green Cape in the Eastern Atlantic. On the Range-70 we have gained the series of measurements with duration about half a year.

The series of measurements were exposed to the band filtration with extraction from the source series of the narrow bandwidth of oscillations of the semidiurnal phase. On the so filtrated series of the oscillations the variability of the amplitudes with the phase about 15 day is visually appreciable. For obtaining the statistical estimate of the fortnightly variability of the amplitudes from the filtered realizations was made the series with the discreteness of 12.4 hours, the length of 360 terms. The spectrum calculated by such series reveals the confidence spike in the vicinity of the 14-daily period.

*Results of the data processing using the non-spectral algorithm.* The Fig. 3.1 shows the part of the "Mezopolygon" map with the buoy stations mapping, which data have been used in the numerical calculations. The measurements of the currents on the horizon of 75 m were analyzed. The numerical calculations of the functional $D(t, c_m, \varphi_m)$ with the different number of the analysable sensor units $N$ have shown, that at usage of the present method the optimal number of the sensor units does not exceed 10-12, and all of them should be arranged in the relative proximity to each other, that is the usage of the given method supposes the analysis of observations of the spatially localized groups of the sensor units - "clusters".

It is explained by the fact, that evidently at the analysis of the observations data of the sensor units arranged far apart from each other it is impossible to use the supposition about the invariance of the parameters of the propagative "wavetrain" in connection with the necessity on such spacing intervals to take into consideration the effects of the horizontal variability, dissipation, dispersive spreading, etc. Thus, the special sizes of the optimal "cluster" of the sensor units are definitely characterizing the spacial scale of the variability of the parameters of the propagative "wavetrain. All the sensor units shown on Fig..3.1, have been divided into the groups (or "clusters") and indicated on the figure by the dashed lines.

The Fig..3.4 presents the results of calculations of the functional $D(t, c, \varphi)$ (the isolines in the plane $c, \varphi$) for the third group of the sensor units. The interval of the average $T$ has made 100 hours. The represented numerical results show, that there are really existing the certain time windows, when the functional $D(t, c, \varphi)$ has one or more rather distinctly expressed stationary (existing at least 50 hours} minimums with respect to variables $c, \varphi$. This fact enables to assert, at first, that the sensor units are really recording the passage of the "wavetrain" with invariable enough parameters, and secondly, that those values of $c, \varphi$, at which there is a stationary minimum $D(t, c, \varphi)$, are precisely enough describing the parameters of the propagating plane wavetrains.

Let's also mark, that the typical minimum stationary values of $D(t, c, \varphi)$ compound 0.12-0.16, and these relatively small values of the stationary minimums testify, that the majority of the sensor units really record the same signal with the corresponding delays record. The results of the wave train analysis show, that the minimum of the functional $D(t, c, \varphi)$ corresponds to the direction of the waves propagating from South-SouthEast, and the most brightly this result is expressed in the end of April, that is completely corresponds to the results of the time-spacial spectrum analysis.

The additional calculations have shown, that just the maximum corresponding to $c = 9.2$ cm/s and $\varphi = 119°$ is stable. The direction is pointed counterclockwise from the null direction to the orient. This maximum is saved at the change of the time of the beginning of integration from 00-00 hours of April, 24th till 00-00 hours of May, 4th. The close values of the maxima were registered and at the analysis of the data from the remaining groups of the sensor units. Study of the stationary minima $D(t, c, \varphi)$ conducted onthe groups of the sensor units arranged along the route of propagation of the discovered "wavetrain" show, that they really registered the same "wavetrain", that enables to determine the whole path of propagation of that "wavetrain" on the whole array of available data.

On Fig. 3.5 the vertical line indicates the averaged latitude value of each group of the sensor units, and the horizontal line indicates the time readings and the time intervals, during which each group of the sensor units observed the wave with the velocity and the propagation direction close to $c = 9.5$ cm/s and $\varphi = 120$. The slanting lines correspond to the movement of the wave train with this velocity. The given data show, that we really are dealt with the "wavetrain", which in series passes through each group of the sensor units. These results enable determination of the whole path of propagation of the "train".

*Interpretation.* The proposed method allows to evaluate the group velocity of the "wavetrain" of the internal waves. The standard methods of the time-spatial analysis give the estimate of the phase velocity of the internal waves, which in the region of "Mezopolygon" is close to the value of 1.5 m\s. For our measurements as the wave train it is considered the increase of the oscillations of the amplitudes of the internal waves caused by the beats of the wave oscillations with the periods of oscillation of 12.4 and 12.0 hours. Duration of such "wavetrain" of oscillations is equal to 14.76 days (see Fig. 3.2). In the case of observation of the internal rising tides on "Mezopolygon" we deal not with the formed stable modes of oscillations, but with a set of the random disturbances propagating along the curved characteristic planes near to the set of the separate sources around each submarine hill. These perturbations are joined each other and shape so complex pattern o the wavetrains, one of which we observe in this experiment. It is known, that the wave vectors of the phase and the group velocities are orthogonally related, at that the wavetrain is propagating more slowly. The ratio between the values the group velocity and the phase velocity is expressed by the following expression:

$$\sqrt{(\omega^2 - f^2)/(N^2 - \omega^2)}.$$

At the value of Brunt-Väisälä frequency equal to $4 \times 10^{-3}$ 1/s, the group velocity compounds about 0.04 of the phase velocity, that is close to 6 cm/s and corresponds to the values gained at calculation of the functional (3.1).

The results of the analysis of the wavetrains have shown, that the minimum of the functional corresponds to the direction of the waves propagating from the South-SouthEast. The most effectively the wavetrain movement is expressed in the end of April. This result coincides with the conclusions of the analysis of the calculations of the time-spatial spectrum gained according to the data of "Mezopolygon", where the main direction of the internal waves movement was determined as propagating from the South. The end of April corresponds to the maximal currents of the barotropic high tide dealt with the syzygial tide period. Thus, the obtained results demonstrate the principle opportunity to use not only the methods of the time-spatial spectrum analysis of the measuements of the accidental hydrophysical fields, but also other methods and approaches. Expansion of the methods used for the measurements data processing enables the most effective way to determine the corresponding characteristics of the measured fields depending on the character of the available data.

**4. Single out of the wave-trains of the internal gravity waves on the background of the heavy interferences in compliance with the results of the in-situ measurements in the Black sea.**

The present section presents the methods of processing of the in-situ measurements of the internal gravity waves conducted in the Black sea, which allow in the conditions of the heavy interferences not only to represent the measured field of the internal gravity waves in the form of the sum of the separate plane wavetrains( at that one of the separated wavetrains may be the wavetrain disturbed by the moving source), but also to determine the characteristics of this "wavetrain". The Fig. 4.1 shows the layout of the experiment (in plan).

For the primary processing of the in-situ measurements data, as well as in sections 2 - 3, it is possible to use the time window allowing to filter off the signals with the spectrum lying outside the typical interval natural for the internal gravity waves. It is interesting to underline, that for the being considered in-situ measurements data the function $E$ describing the slow variations of the corresponding parameters ( the temperature, the horizontal components of velocity) calculated in compliance with the offered in sections 2 - 3 algorithms is much more by the amplitude, than the function $S$, which describes the internal gravity waves themselves. It testifies that the rather low-frequency components are mainly dominating in the analysable data with the characteristic period of variation equal to one hour and more. The Fig. 4.2 represents the typical record of one of sensor units (function $S$, the meridaonal velocity), its duration - three days. As it is evident from the Figure, the measured field represents the complex enough picture of the quasi-harmonic oscillations. At that it is impossible to determine uniquely the presence of the expectable (according to the theoretical calculations) signal from the moving source.

Statement of the problem considered in this section about definition of the "wavetrain" excited by the movement of the source will consist in the following: determination of thea time of iarrival of the " wave train" (the point of observation - any sensor unit), the direction and velocity of propagation of this "wavetrain" , duration of the "wavetrain", and also the distance up to the trajectory of the source movement (see Fig.4.1 - BC line) - the traverse distance.

The velocity and the direction of propagation of the "wavetrain" can be determined from the overdetermined system of equations (2.2), however, as the conducted numerical calculations have shown, because of the high noise level of the initial in-situ measurements data, simultaneously there are several maxima of the correlation function $K_{ij}(t,\tau)$, that does allow unambiguously to determine the time of delay of the separate "wavetrain" and so we used the method offered in the section 3. The conducted numerical calculations for a time span, during which there are simultaneously the readings of no less than from three sensor units, have shown, that mainly there is the possibility to determine three versions of existence of the minimums $\tilde{D}$ of the functional $D(t,c,\varphi)$:

1. The $\tilde{D}$ values are non-stationary (the values $c,\varphi$ determining the value of a minimum depend on a time), the typical values of $\tilde{D}$ lie within the limits of 0.7-0.8, it means, that the majority of sensor units do not register one and the same signal (Fig.4.3).

2. There is one stationary minimum $\tilde{D}_1$. At that the typical value of $\tilde{D}_1 = 0.3$-$0.4$. The Fig.4.3,a shows the isolines of the functional $D(t,c,\varphi)$ ( the meridaonal velocity, the first day), value of t is equal to 14 hours 30 minutes, the minimum $\tilde{D}_1 = 0.44$ is reached in the point $c_1 = 0.13$ m\s, $\varphi_1 = 108°$. The values of $c,\varphi$ determining the $D(t,c,\varphi)$ remain practically invariable within 1.5-2.5 hours.

3. There are two stationary minima $\tilde{D}_1$, $\tilde{D}_2$, at that their typical values lie within the limits of $\tilde{D}_1 = 0.3$-$0.4$, $\tilde{D}_2 = 0.5$-$0.6$. The Fig,. 4.3,b there are isolines $D(t,c,\varphi)$ (the meridaonal velocity, the first day), the value of t is equal to 21 hours 45 minutes, the minimum points: $\tilde{D}_1 = 0.39$, $c_1 = 0.08$ m/s, $\varphi_1 = 276°$, $\tilde{D}_2 = 0.51$, $c_2 = 0.13$ m/s, $\varphi_2 = 163°$. As well as for the event of one stationary minimum, the values of $c_{1,2}$, $\varphi_{1,2}$ remain invariable within 1-2 hours.

Vectors of the velocity of propagation of the internal gravity waves are expressed as
$V(t) = (c(t)\cos\varphi(t), c(t)\sin\varphi(t))$, where the values $c(t), \varphi(t)$ determine the minimum $D(t,c,\varphi)$ in the moment t and are shown on the Fig. 4.4, their beginning – 13 o'clock of the first day (the solid vector shows the values of $c,\varphi$, which are determined from the absolute minimum $D(t,c,\varphi)$; the dashed vector shows the values of $c,\varphi$, which are determined from the second minimum $D(t,c,\varphi)$, ( if it exists). From the presented results one can see, that there are time intervals, when at the given tine the sensor units record only one "wavetrain", at the same time there are time intervals, when the sensor units record two existing wavetrains. In addition the sensor units rather often can register non-stationary chaotic wave motion.

On the Fig. 4.4 the points $T_{1,2}$ mark the times of the wavetrain" incoming from the moving source calculated on the basis of the experiment geometry. At that the direction of the wave propagation lies within the limits of 110° - 130°, and, as it is quite evident, the time of existence of the stable "wavetrain", as well as the direction of its propagation with the sufficient degree of accuracy coincide with the calculated time of arrival the signal and the angle of its propagation. The velocity of the "wavetrain" propagation has coincided with the maximal group velocity of the first mode of the internal gravity waves equal to l 0.13 m/s, which one determines the moment of arrival of the wave front excited by the moving source.

It is evident, that the nonstationary behavior of the medium and dependence of its parameters on the horizontal coordinates results in the change of the media amplitude-phase characteristics of the fields of the internal gravity waves, and, accordingly, to the change of the time of arrival of the "wavetrain" from the source, as well as the direction of its propagation. As one can see from the numerical calculation conducted according to the methods, set forth in [1-8], for the hydrology of the given experiment the taking into consideration of the nonstationarity of the medium and the horizontal nonuniformity of the field density (the sea depth is practically constant) result in the corrections of the propagation direction of the "wavetrain" of the order of 15° and of the time moment of incoming of the "wavetrain" arrival from the moving source of the order of 20 minutes.

Now we shall go to consideration of the problem of single out the wavetrain caused by the moving source among other stable "wavetrains". For determination of the moment of arrival and the characteristics of the wavetrain disturbed by the moveing source we shall take advantage of the results of the linear theory of the internal gravity waves. As it is known, the temperature T and a horizontal speed **u** (these parameters were measured in the given experiment) are coupled to the elevation η by the ratio

$$T = T^*\eta, \qquad \Delta \mathbf{u} + \frac{\partial^2 \nabla}{\partial t\, \partial z}\eta = 0, \qquad \nabla = \left(\frac{\partial}{\partial x}, \frac{\partial}{\partial y}\right)$$

where $T^*$ is the average temperature gradient in depth.

Then, as follows from the results of [1-8] far from the moving disturbing body, which we shall approximate the dipole, the field of the separate mode of the internal gravity waves is described by the asymptotic form (2.2.21) (the far asymptotics of the shallow sea)

$$T = T_0\, \psi(z)\, Ai(\alpha(t - t_0))$$
$$\mathbf{u} = \mathbf{n}\, u_0\, \psi'(z)\, Ai(\alpha(t - t_0)) \tag{4.1}$$

where $T_0, u_0$ are the amplitude multipliers, $\psi(z), \psi'(z)$ are the vertical profiles of the corresponding eigenfunction and its derivative, $Ai(x)$ is Airy function, $\alpha$ is the parameter depending on first two factors of decomposition of the dispersion curve in zero and the traverse distance, $t_0$ is the moment of arrival of the "wavetrain", **n** is the ort direction of its propagation. Let's call the expressions (4.1) as the expected signal. Further we shall consider the integral

$$Z_i(t,\tau) = \int_{t-b/2}^{t+b/2} S_i(\rho)\, \Phi(\rho + \tau)\, d\rho$$

where $S_i$ is the reading of the i-th sensor unit, $\Phi$ is the expected signal on the given sensor unit. If $\tau$ concurs the due time of arrival of the "wavetrain" on the given sensor unit, then it is possible to assume, that the signal recorded on this sensor unit will be similar to the expected signal, then at the given $\tau$ function $Z_i(t,\tau)$ will have the maximum. However, the numerical calculations show, the functions $Z_i(t,\tau)$ have the significant number of the local maxima, to determine among which the maximum corresponding to the moment of arrival of the "wavetrain" from the moving source, unambiguously does not seem possible.

Therefore, further we shall represent the expected signal in the following form: $\Phi = \Phi_1 + \Phi_2$, where $\Phi_1$ corresponds to the first two half-waves of then expected signal (the beginning of the signal), $\Phi_2$ describes the following oscillations, $\Phi(t)$ (the end of the signal), and we shall consider the function $\widetilde{P}_i$, which looks like

$$\widetilde{P}_i = \begin{cases} P_i^1 P_i^2, & P_i^1 P_i^2 > 0 \\ 0, & P_i^1 P_i^2 < 0 \end{cases}$$

$$P_i^{1,2} = \int_{t-b/2}^{t+b/2} S_i(\rho)\, \Phi_{1,2}(\rho + \tau)\, d\rho$$

Usage of the function $\widetilde{P}_i$ allows to take into consideration those moments of time, when the analyzable signal coincide both with the beginning and with the end of the expected signal. Let' assume further, that the expected signal is recorded by L sensor units, and we shall calculate the sums of the following kind

$$\sigma_k = \left(\frac{1}{C_L^k} \sum_{i_1 \neq i_2 \neq \Lambda\, \neq i_k} \widetilde{P}_{i_1} \widetilde{P}_{i_2} \Lambda\, \widetilde{P}_{i_k}\right)^{1/k}$$

If all the sensor units at one and the same time record the signal close to the expected signal, then $\sigma_k \neq 0$ at all values of $k$. At the same time the accidental local maxima of the functions $\tilde{P}_i$, which are not allowing for one sensor unit unambiguously to determine the moment of arrival of the "wavetrain" from the moving source, will convert into zero at the values of $k$ close to the total number of the sensor units $L$. We shall note also, that failure of some sensor units can lead to vanishing of the true maxima of $\tilde{P}_i$. In this case it is necessary to consider the functions $\sigma_k$ with the values of $k$ smaller, than $L$. The Fig. 4.5 presents the results of calculations of the functions $\sigma_k$ (meridaonal velocity, the first day), $k = 1,2,3,4$, the points of time of arrival of the "wavetrain" from the moving source are shown by the point $T_{1,2}$ calculated from the geometry of the experiment.

For determination not only the moment of time of arrival of the "wavetrain", but also the traverse distance $y$, which, generally speaking, is not known, it is necessary to scan the values of the parameter $\alpha$, since the previous numerical calculations were conducted at the fixed value $\alpha = \tilde{\alpha}$, determined using the known value of $y$: $\tilde{\alpha} = (q/3\beta y)^{1/3}$, where $q, \beta$ are the first two factors of the expansion of the dispersion curve $\mu(\nu)$ [1-8]. The conducted numerical calculationsshow, that for the values of $\alpha$ corresponding to the true value of $y$, the time of arrival of the "wavetrain" is determined unambiguously, and practically the false maxima of the functions $\sigma_k$ ($k = 3,4$).. It is necessary to underline, that such character of the behavior of $\sigma_k$ takes place for all the channels of the measurements (two components of the horizontal velocity, temperature).

Deviation of the parameter $\alpha$ from the value of $\tilde{\alpha}$ results in the fact, that the functions $\sigma_k$ will have the false maxima with the amplitude, as a rule no less than the maximum value of $\sigma_k$ at $\alpha = \tilde{\alpha}$. At that we shall note, that for each channel of measurements it is typical to have its own position of these maxima. Therefore the criterion for determination of the true value of the parameter $\alpha$ (the traverse distance - $y$) it is possible to formulate as follows: at the values of $\alpha$ corresponding to the true value of $y$, the function $\sigma_k$ (the value $k$ are close to the total number of the sensor units) should have the local maxima at one and the same time for the majority of channels of measurements.

Thus, the conducted analysis of the in-situ measurements data with the purpose to single out the "wavetrain" demonstrates the feasibility t apply the linear theory of the internal gravity waves, in particular, the asymptotic representation of the far field of the internal gravity waves, for the analysis of the data of the in-situ observation of these waves, and the represented analysis algorithms allow to single out the "wavetrain" caused by the sources of disturbancesing bodies among other wavetrains and the random interferences.

**Figures**

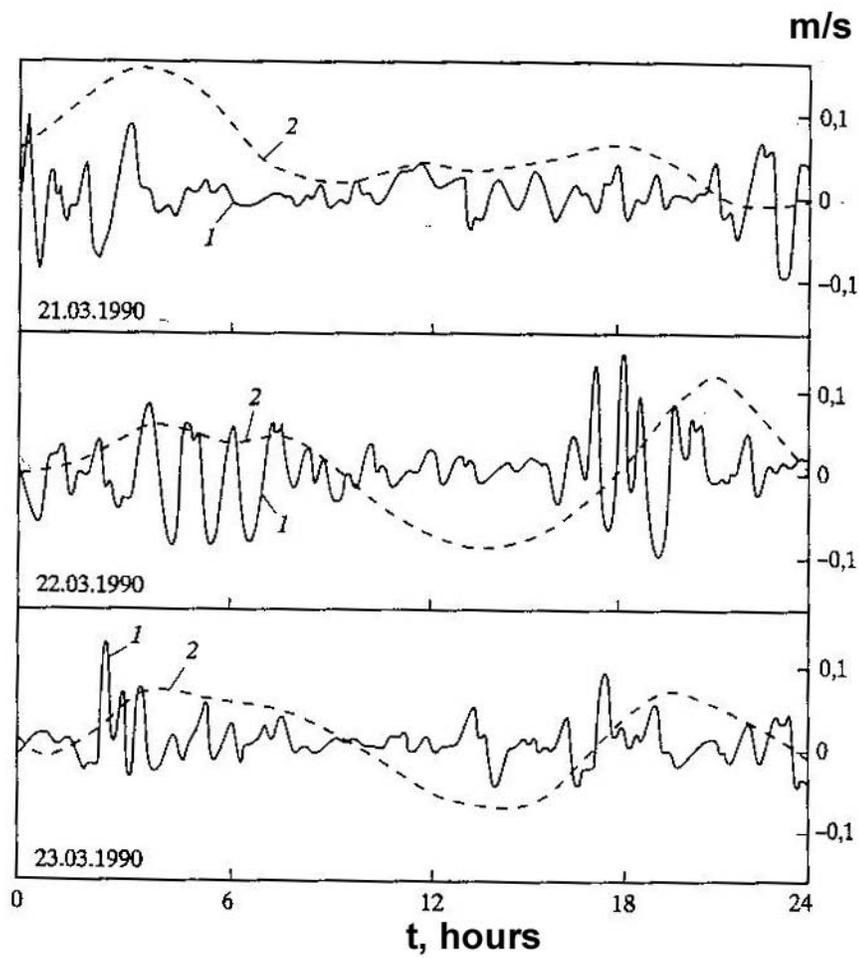

Fig. 2.1    The three days record of the meridional velocity.

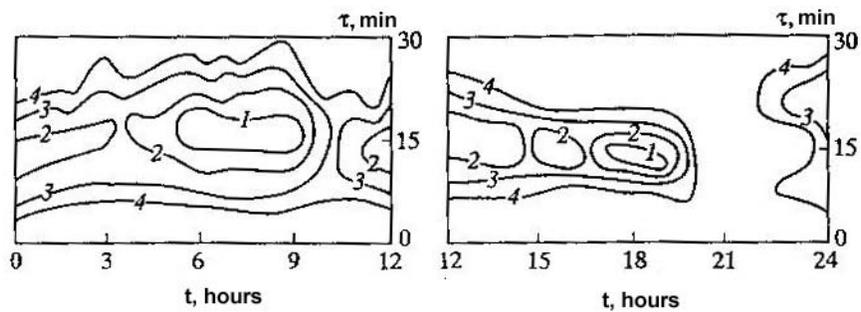

Fig. 2.2 The isolines of the correlation function $K_{ij}(t,\tau)$ : 1 - 0.9 $K^*$, 2 - 0.8 $K^*$, 3 - 0.7 $K^*$, 4 - 0.6 $K^*$, where $K^*$ - the maximum value $K_{ij}(t,\tau)$.

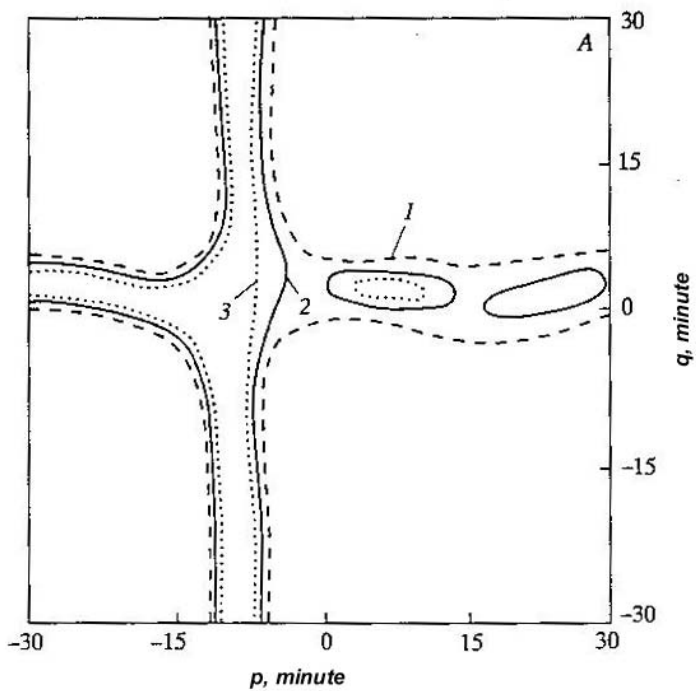

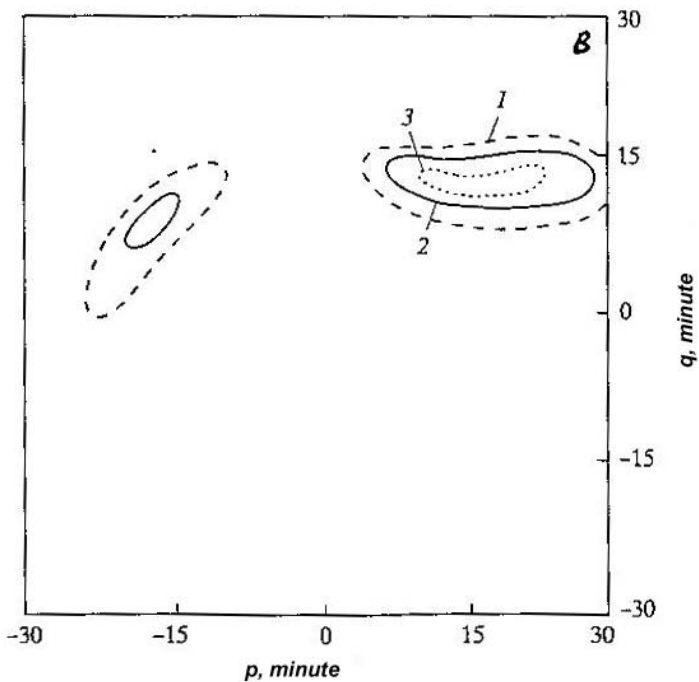

Fig. 2.3  The isolines of the functional $F(t,u,v)$: values of levels 1.3 $F^*$ (the dashed line), 1.2 $F^*$ (the solid line), 1.1 $F^*$ (the dotted line), where $F^*$ - the minimal value $F(t,u,v)$.

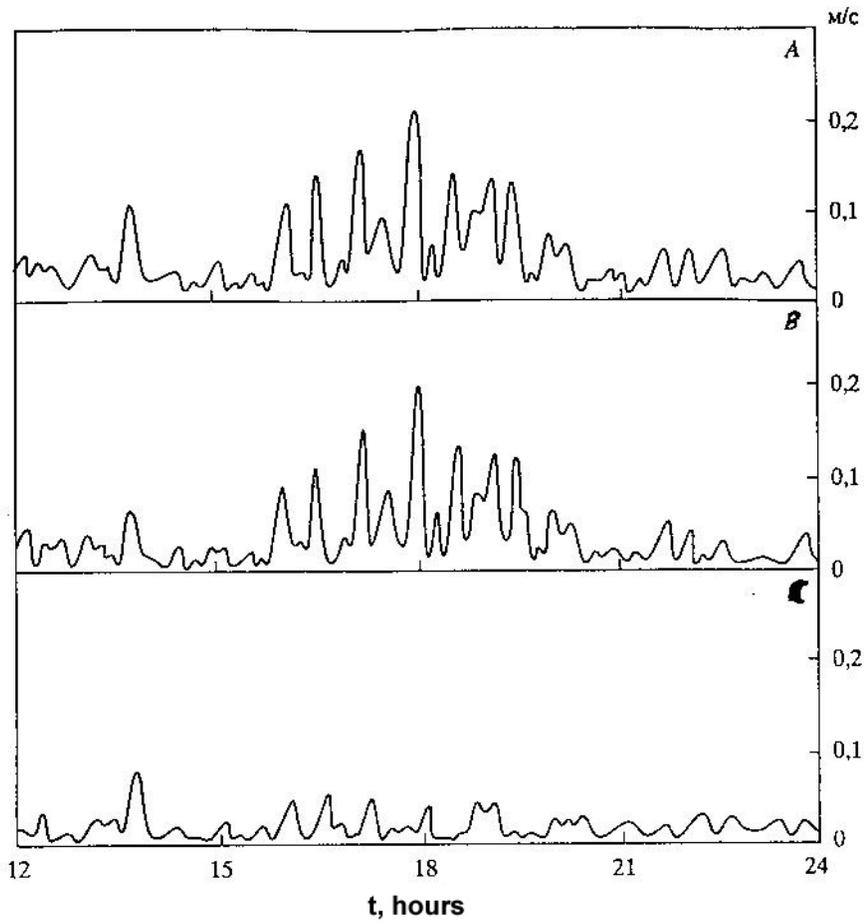

Fig. 2.4  The internal waves horizontal velocity expansion with respect to the internal waves propagation direction.

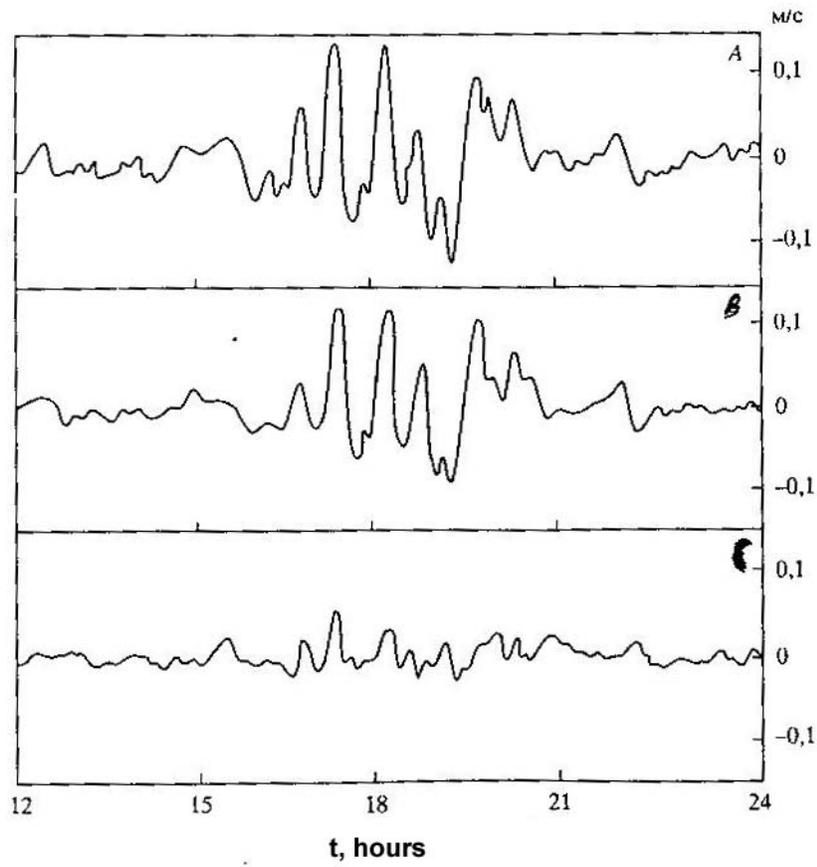

Fig. 2.5   Definition of the plane wave form.

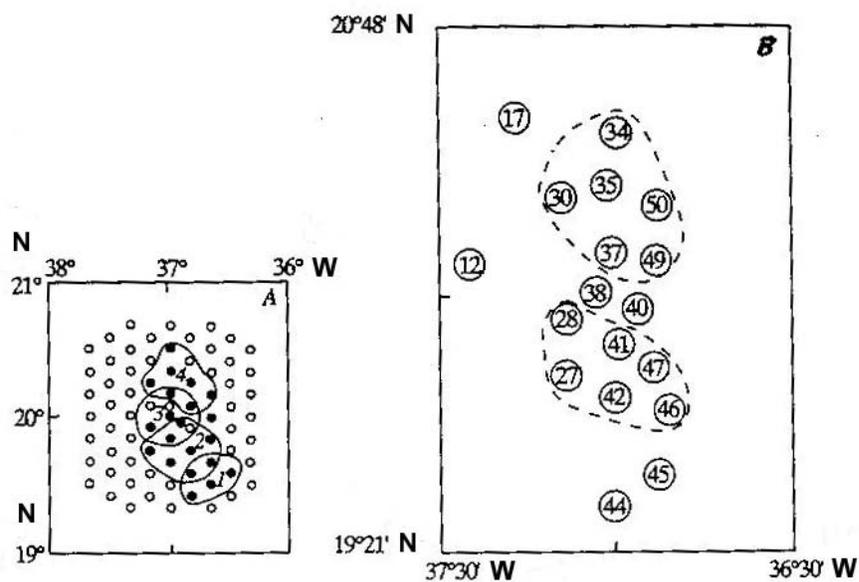

Fig. 3.1 Arrangement of the buoy stations on 75 m horizon: A - the black color marks the buoys, which data were analyzed; the groups of detectors of the buoy stations (1,2,3,4) are encircled by the fine lines; B - the detailed fragment of "Mesopolygon" with indication of the buoy stations numbers.

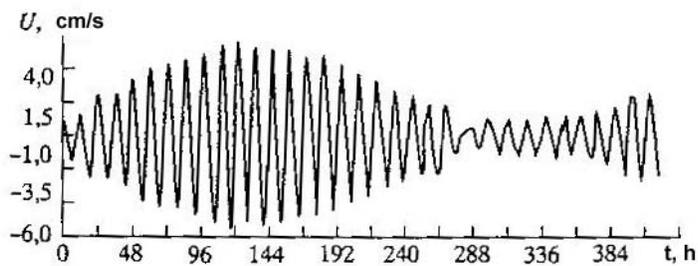

Fig. 3.2 Example of the package of the tidal internal waves recorded in the field of the current velocity by the buoy station 50.

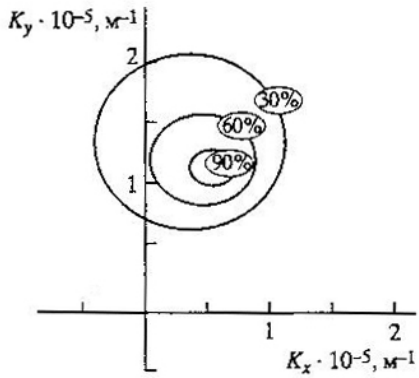

Fig.3.3 The space-time spectrum of the temperature fluctuatings in 200 m horizon calculated on the basis of the measurements of the buoy stations shown in Fig.5.3.1.

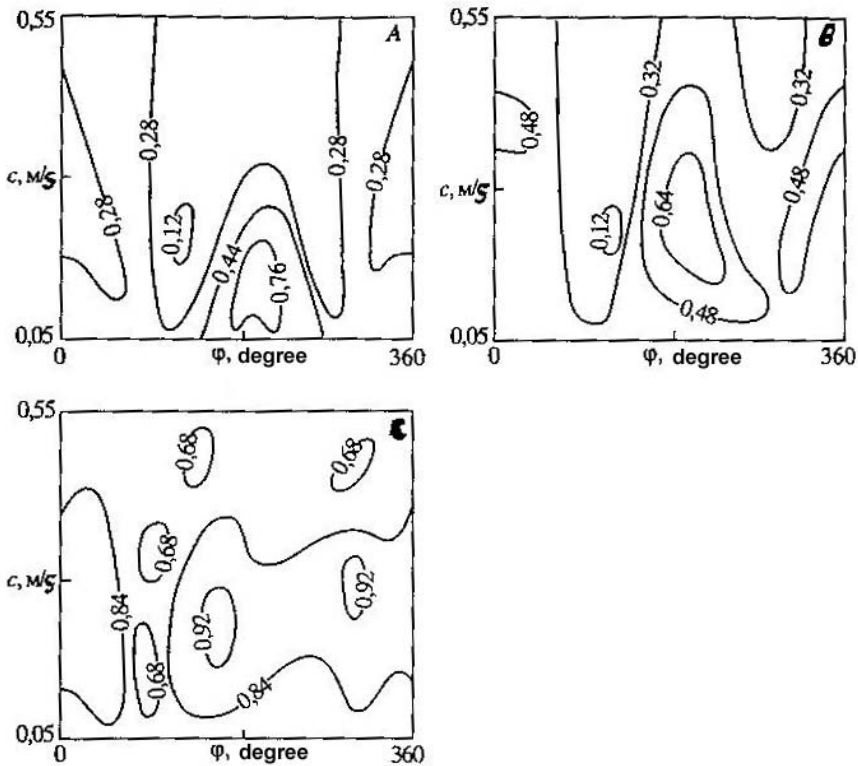

Fig. 3.4 Isolines of the functional $D(t, c, \varphi)$ for detectors 27, 28, 41, 42, 46, 47. Time of the beginning of integration: A - on April 22-nd, 1985, 00-00 o'clock, B - on April 26-th, 1985г., 00-00 o'clock, C - on April 30th, 1985, 00-00 o'clock. The direction of propagation of the wave train $\varphi$ is marked in degrees along the horizontal line (counter-clockwise from the direction to the east), along the vertical line – the group velocity $c$.

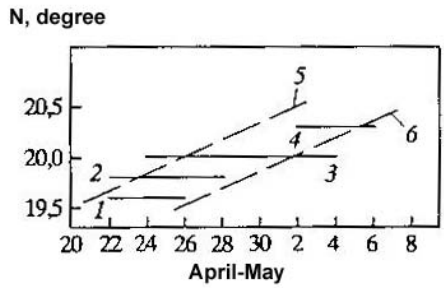

Fig. 3.5 Propagation of the internal waves train with the group velocity about 10 cm/s and the direction of $120^0$. The vertical line demonstrates the mean latitude of an arrangement of the groups of detectors; the horizontal line shows the time marks. The direct horizontal lines indicate the intervals of time, during which each group of the detectors watched the wave train. The slant dashed lines characterize the velocity of propagation of the wave train.

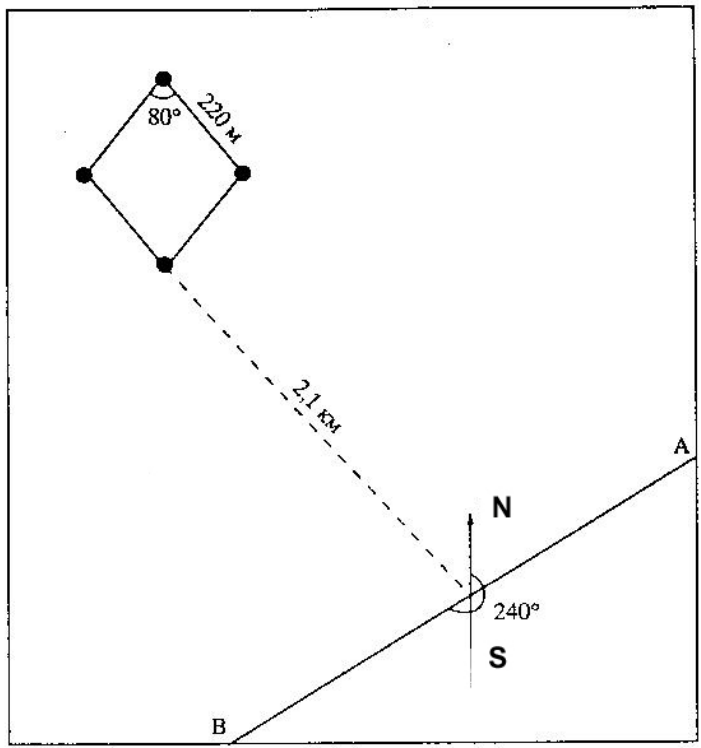

Fig. 4.1 The scheme of the experiment of measurements of the internal gravity waves in the Black sea.

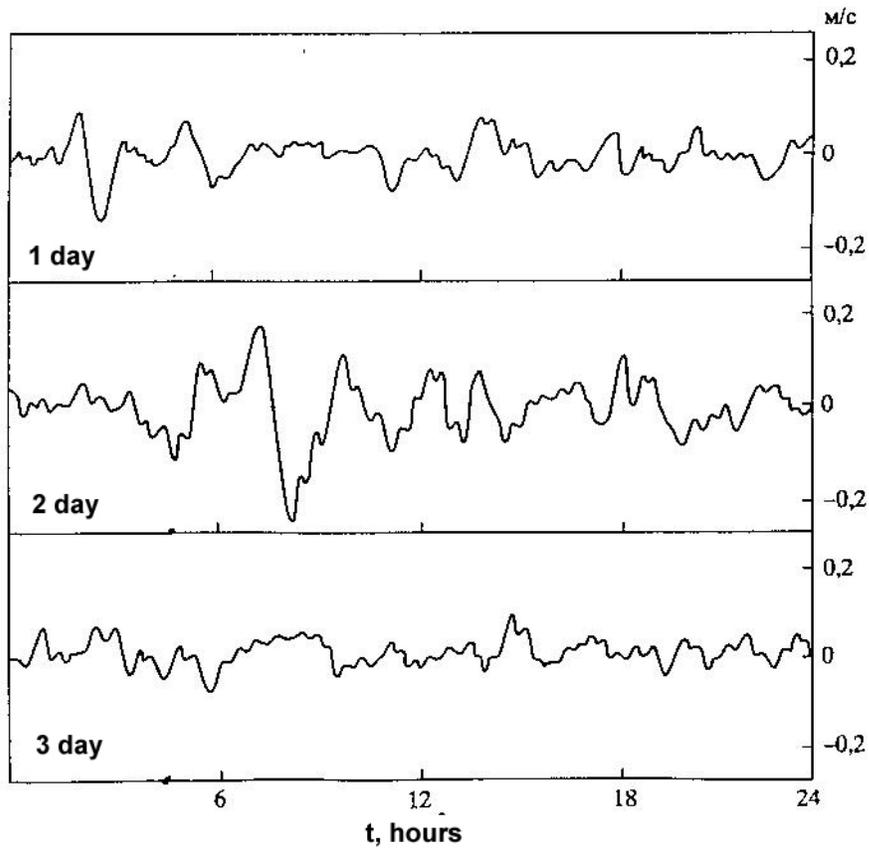

Fig. 4.2  3 days records of the meridional velocity.

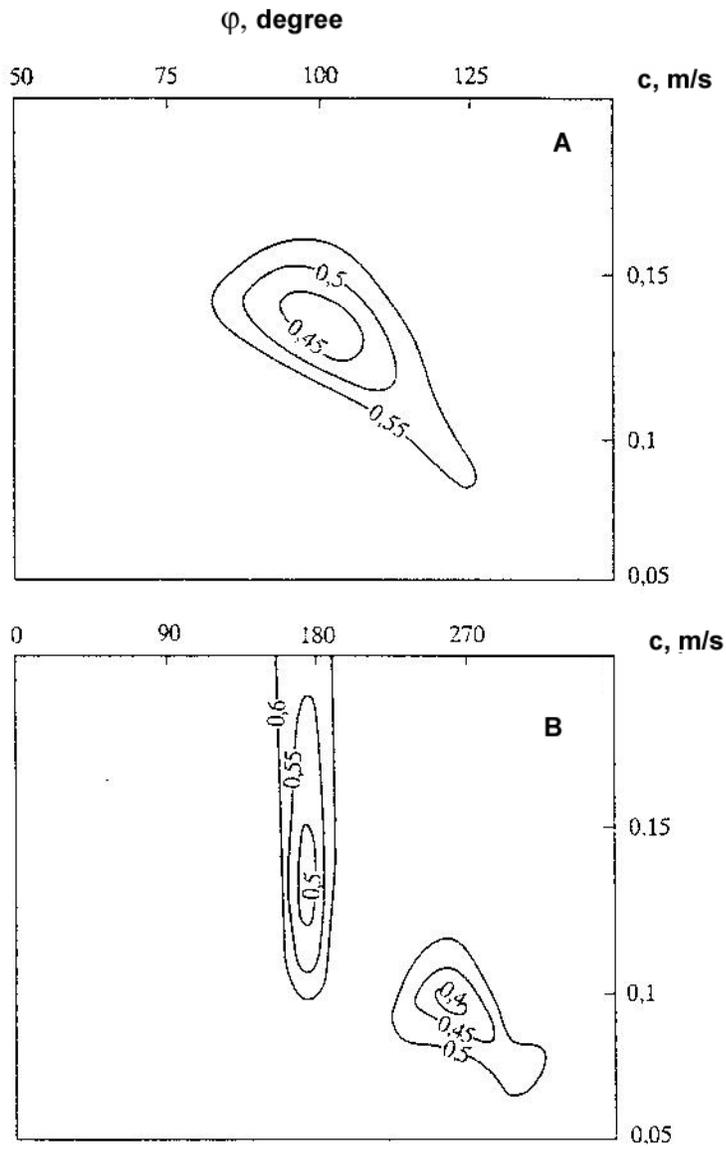

Fig. 4.3    The isolines of the functional $D(t, c, \varphi)$.

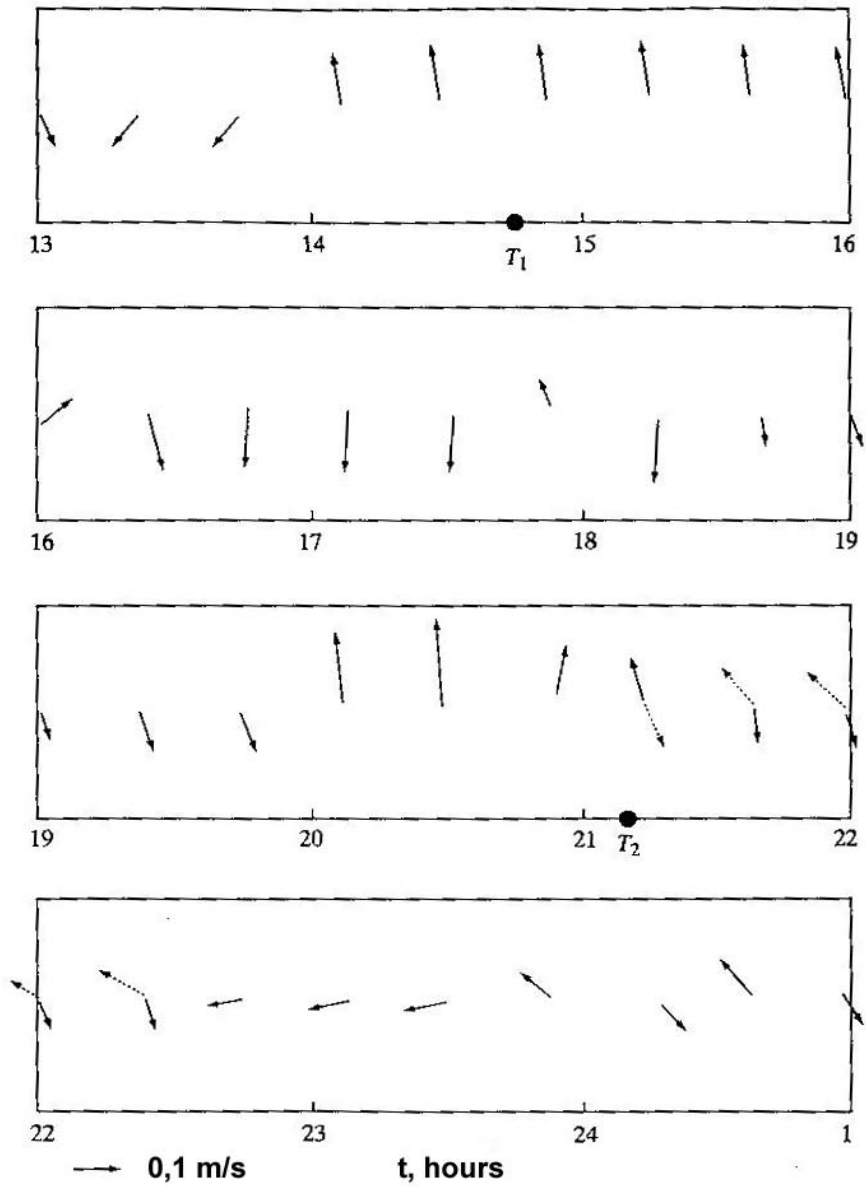

Fig. 4.4 Vectors of the wave trains propagation velocities.

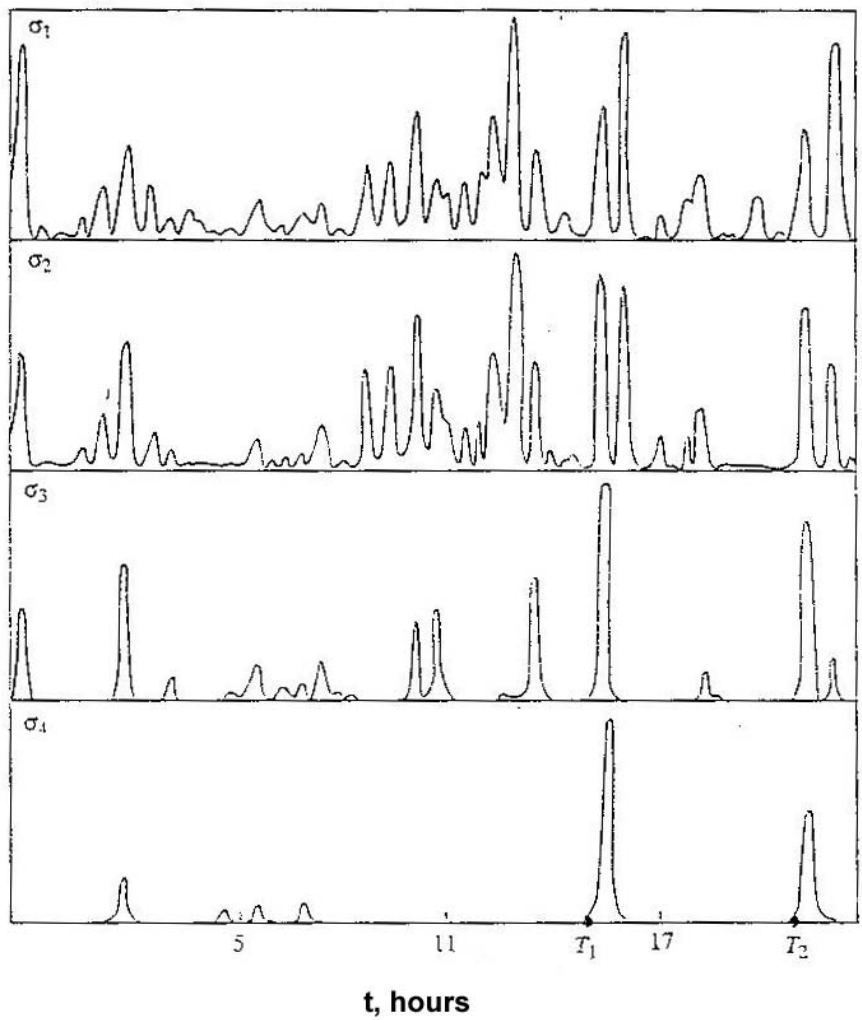

Fig. 4.5  Identification of the wave trains against the background of the great disturbances - the functions $\sigma_k$ (k=1,2,3,4).